\pdfoutput=1 
\documentclass{JINST}

\usepackage{subfigure}
\usepackage{epsfig}
\usepackage{amsmath}
\usepackage{multirow}
\usepackage{caption}
\usepackage{rotating}

\title{Material Screening with HPGe Counting Station for PandaX Experiment}

\author{
Xuming Wang$^a$,
Xun Chen$^a$,
Changbo Fu$^a$\thanks{Corresponding author, cbfu@sjtu.edu.cn.},
Xiangdong Ji$^{a,c}$,
Xiang Liu$^a$\thanks{Now at Max-Planck-Institut f\"ur Physik, Germany.},
Yajun Mao$^b$, 
Hongwei Wang$^d$,
Siguang Wang$^b$,
Pengwei Xie$^a$,
Tao Zhang$^a$\\
\llap{$^a$}{\it INPAC}, Department of Physics and Astronomy, Shanghai Jiao Tong University, 200240, Shanghai, China\\
\llap{$^b$}School of Physics, Peking University, 100871, Beijing, China\\
\llap{$^c$}Department of Physics, University of Maryland, College Park, MD, 20742, USA\\
\llap{$^d$}Shanghai Institute of Applied Physics, Chinese Academy of Sciences, 201800, Shanghai, China\\
}

\abstract{
A gamma counting station based on high-purity germanium (HPGe) detector was set up for the material screening of the PandaX dark matter experiments in the China Jinping Underground Laboratory.
Low background gamma rate of 2.6 counts/min within the energy range of 20 to 2700~keV is achieved due to the well-designed passive shield.
The sentivities of the HPGe detetector reach mBq/kg level for isotopes like K, U, Th, and even better for Co and Cs, resulted from the low-background rate and the high relative
detection efficiency of $175\%$. The structure and performance of the counting station are described in this article.
Detailed counting results for the radioactivity in materials used by the PandaX dark-matter experiment are presented.
The upgrading plan of the counting station is also discussed.
}


\begin{document}

\section{Introduction}
\label{sec:intro}
The PandaX\cite{Cao:2014jsa} project, located in the China Jinping Underground Laboratory (CJPL)\cite{1742-6596-203-1-012028}, was initiated for searching rare physical events in particle and astrophysical physics with xenon detectors. 
Its first and second stage (PandaX-I\cite{Xiao:2015psa} and PandaX-II\cite{Tan:2016diz}) experiments are for dark matter search\cite{Akerib:2015rjg, Baudis:2012zs}. 
Low-background rate is required for the detection of rare dark matter signals, thus the low radioactive background controlling is one of the most critical issues. 
The radioactivity in the materials used for the detectors must be sufficiently low to ensure a good sensitivity for physical signals. 
Measurements of the material backgrounds are the first step for the constructions of the detectors. 
Various techniques have been developed to measure the low-level radioactivities, such as neutron activation analysis (NAA), inductively coupled mass spectrometry (ICP-MS), and high purity germanium (HPGe) gamma counting. 
Because of its wide energy window (several keV to tens of MeV) and high energy resolution, HPGe detector can be used to measure the most radioactive isotopes, such as $^{60}$Co, ${137}$Cs, $^{40}$K and the decay products from the uranium and thorium chains. 
HPGe counting uses a nondestructive method, so it is not only useful in screening samples, but also in assaying of finished products.

A HPGe counting station has been set up for the material screening of PandaX-I and PandaX-II in CJPL. 
In this paper, we describe the construction of the counting station in Sec.~\ref{sec:overview} and discuss its performance in Sec.~\ref{sec:performance}. 
The calibration and material assay process for the PandaX experiment is described in Sec.~\ref{sec:calib_counting}, and the results are presented in Sec.~\ref{sec:result}. 
A summary is given in the last section, together with an upgrading plan of the counting system.

\section{Overview of the Counting Station}
\label{sec:overview}
The PandaX gamma counting station is comprised of a HPGe detector and a passive shielding system.

The HPGe detector was fabricated by Ortec, customarily designed for the purpose of radiation counting. It consists of a P-type coaxial HPGe crystal, a cooling system and a pre-amplifier electronics. The sensitive mass of the crystal is 3.69~kg, resulting in a relative detection efficiency of 175\% (to a standard NaI scintillator detector). The large sensitive mass makes it very effective for the detection of photons within the energy range of 0.01 to 20 MeV. The specification of the detector provided by the manufacturer is given in Tab.~\ref{tab:detector-specification}.

\begin{table}
\center
\begin{tabular}{ll}
\hline \hline
 Manufacturer                 &ORTEC                             \\
 Detector model No.           &GEM-MX94100-LB-C-HJ-S             \\ \hline
{\bf Performance}  &                                             \\
 Relative efficiency          &175$\%$                           \\
 Threshold                    &10 keV                            \\
 FWHM at 1.33 MeV, $^{60}$Co  &2.3 keV                           \\
 FWHM at 122 keV, $^{57}$Co   &1300 eV                           \\
 FWHM at 14.4 keV, $^{57}$Co  &1280 eV                           \\ \hline
{\bf Crystal}       &                                            \\
 Sensitive crystal mass       &3.69 kg                           \\
 Crystal diameter             &93.8 mm                           \\
 Crystal length               &103.7 mm                          \\
 Hole diameter                &11.2 mm                           \\
 Hole depth                   &89.9 mm                           \\
 Dead layer thickness         &0.015 - 0.7 mm                    \\ \hline
{\bf Cryostat}    &                                              \\
 End cap diameter             &108 mm                            \\
 End cap thickness            &0.9 mm                            \\
 End cap to crystal           &5 mm                              \\
 End cap material             &Carbon fiber                      \\ \hline
\hline
\end{tabular}
\caption{The specifications of the HPGe detector used in the gamma counting station .}
\label{tab:detector-specification}
\end{table}

The main purpose of the passive shielding system is to reduce the background-radiation level resulting from the ambient environment. Since the cosmic muon flux inside CJPL is very low\cite{Yu-Cheng:2013iaa}, a cosmic muon veto is unnecessary.

\begin{figure}
\begin{center}
\includegraphics[width=0.9\textwidth]{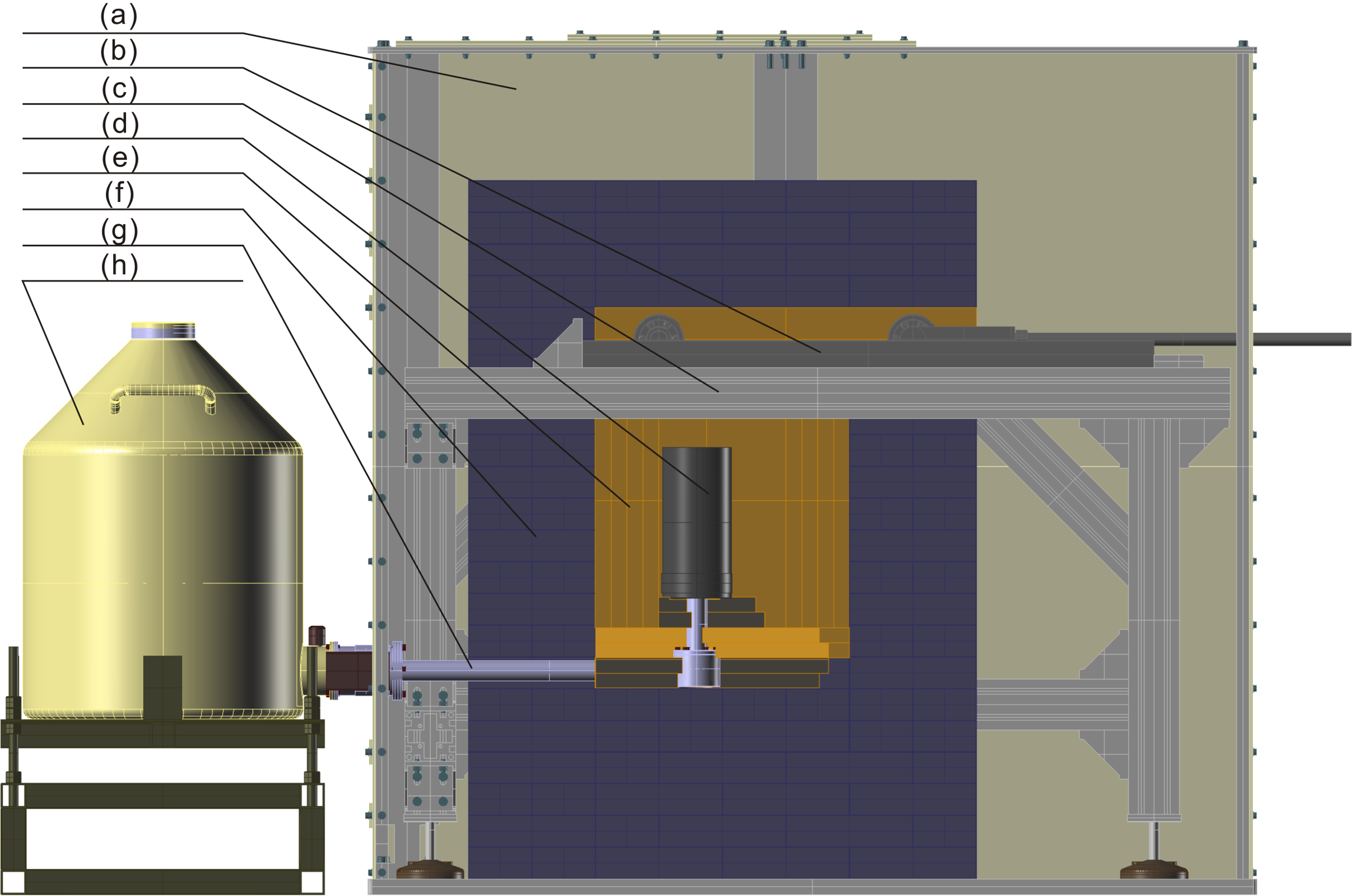}
\end{center}
\caption{
Schematic view of the PandaX gamma counting station.
(a) The air-tight acrylic shell;
(b) The guide rail of aluminum alloy;
(c) The supporting structure aluminum alloy;
(d) Carbon fiber cover with HPGe crystal;
(e) Copper chamber;
(f) Lead shield;
(g) Cooling finger with cable;
(h) Dewar for the storage of nitrogen.
}
\label{fig:side-view}
\end{figure}

A schematic view of the counting station is shown in Fig.~\ref{fig:side-view}. The HPGe crystal together with a cover of carbon fiber is located in a chamber made of high-purity oxygen free (OFHC) copper at the center of the shielding system (part d). The dimension of the chamber is $20\times20\times35$~cm$^{3}$ and the thickness of the chamber Cu wall is 10~cm. Samples to be counted will be placed in the chamber during operation, and the available space is 11.92~L. An OFHC copper stand with a height of 5~cm is placed at the bottom the chamber to support the HPGe crystal. The bottom of the chamber and the stand were designed to match the curvature of the HPGe cryostat finger seamlessly, providing protection to the finger and background shielding. The chamber is enclosed by a lead shield, which is made of more than 500 lead bricks, and each brick has a dimension of $20\times10\times5$~cm$^{3}$. The bottom of the lead shield has 30~cm in thinness and the other faces are 20~cm thick, respectively. The top of the lead shield and the copper chamber are fixed on a guiding rail of aluminum alloy, so they can move along the rail horizontally, enabling the physical access to the chamber. The guiding rail is supported by an aluminum-alloy structure. An air-tight acrylic shell covers the whole shielding system to prevent Rn leaking from the air. The cooling system for the detector consists a liquid nitrogen dewar with a capacity of 30~L, a cooling finger with a length of 63~cm. Since the dewar is placed outside the shielding system, the long cooling finger ensures enough shielding space for the HPGe crystal.

The lead and OFHC copper used for the construction of the shielding system were screened by the present counting station and the HPGe counting facility\cite{ARPESELLA1996991} at Laboratori Nazionali del Gran Sasso (LNGS), respectively. The counting results of the main radioactive isotopes are summarized in Tab.~\ref{tab:radiation_copper_lead}.
They provide one of the important constrains on the sensitivity of this detector.

\begin{table}
\center
\begin{tabular}{ccc}   \hline
\hline
Material & Radioactive Isotope & Radioactivity [mBq/kg]      \\ \hline
\multirow{9}{*}{Copper}
         & $^{60}$Co       & 0.20$\pm$0.09                   \\ \cline{2-3}
         & $^{56}$Co       & 0.20$\pm$0.07                   \\ \cline{2-3}
         & $^{58}$Co       & 1.2$\pm$0.2                     \\ \cline{2-3}
         & $^{40}$K        & 4$\pm$1                         \\ \cline{2-3}
         & $^{137}$Cs      & $<$0.16                         \\ \cline{2-3}
         & $^{54}$Mn       & 0.19$\pm$0.08                   \\ \cline{2-3}
         & $^{238}$U       & $<$0.38                         \\ \cline{2-3}
         & $^{232}$Th      & $<$0.51                         \\ \cline{2-3}
         & $^{235}$U       & $<$0.86                         \\ \hline
  Lead   & $^{210}$Pb      & 300                             \\ \hline
\hline
\end{tabular}
\caption{The counting results of the radioactive isotopes in the copper and lead used for the construction of the shielding system.}
\label{tab:radiation_copper_lead}
\end{table}

Rn is purged by flushing the chamber continuously with boiled-off nitrogen gas. The acrylic shell is not entirely gas-tight, therefore after opening the chamber and
exchanging samples, it usually takes about 12 hours for the Rn to drop to its lowest level. Assume all background in the measurement
come from the Rn, its level inside the chamber is calculated to about 1.5~Bq/m$^3$, which is two-order of magnitude lower than without purging.

A detailed analysis of the background sources along with a full Monte Carlo simulation is discussed in Appendix.

\begin{figure}
\centering
\includegraphics[width=0.8\textwidth]{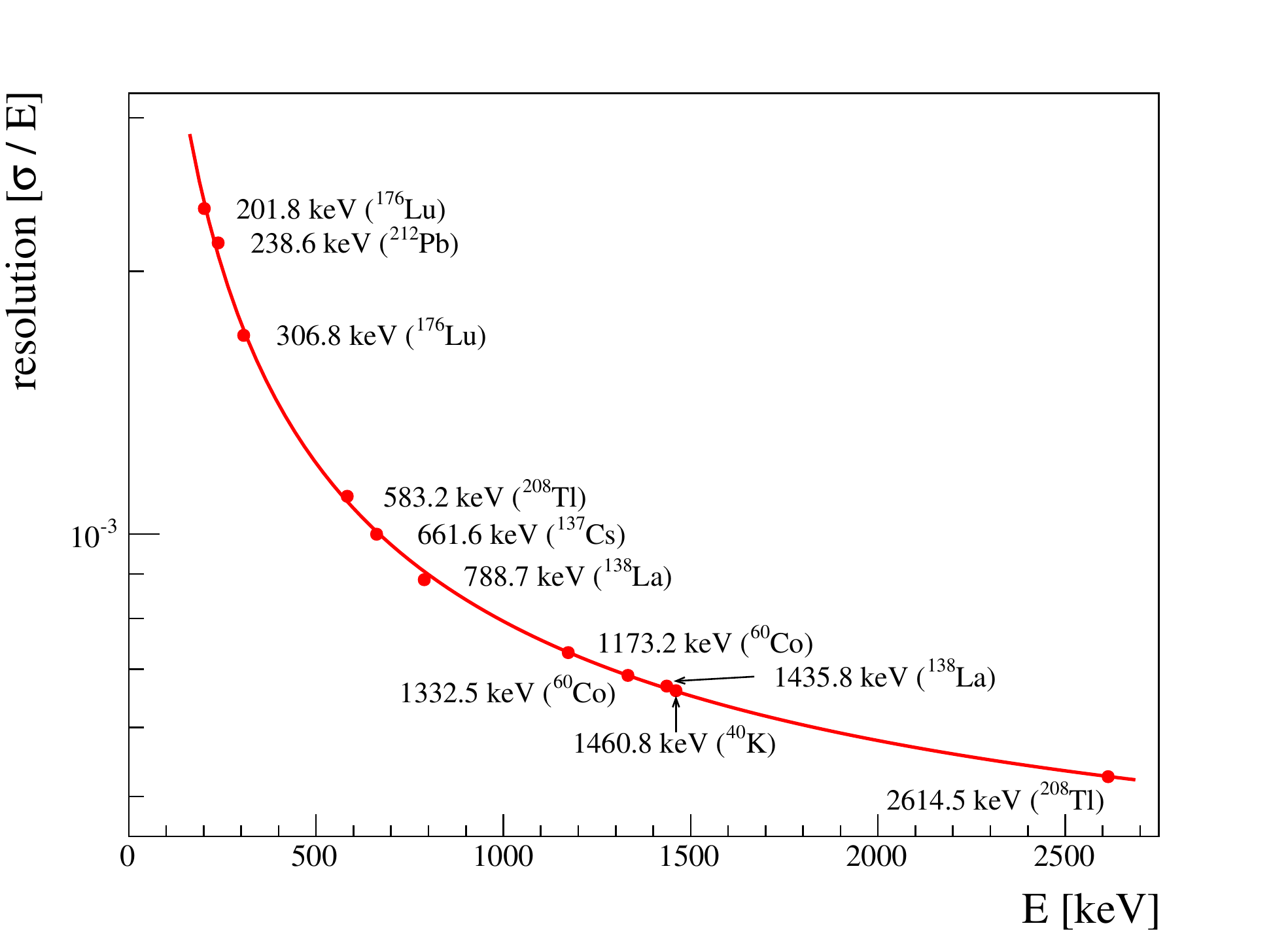}
\caption{The calibrated resolution of PandaX counting station at different energies. The red solid line is the fitted relation between $\sigma$ and energy $E$.}
\label{fig:res_calib}
\end{figure}

The energy resolutions $\sigma/E$ of the detector are calibrated with different characteristic lines from the standard sources of $^{137}$Cs, $^{60}$Co and natural mineral powder of La$_2$O$_3$, where $\sigma$ is the measured variance for an energy. The isotopes of $^{176}$Lu, $^{212}$Pb, $^{208}$Tl, $^{137}$Cs, $^{60}$Co, $^{138}$La and $^{40}$K have been seen from the gamma spectrum. The calibration results are shown in Fig.~\ref{fig:res_calib}. The relation between the variance $\sigma$ and the energy $E$ (in keV) can be expressed by a polynomial
\begin{equation}
\label{eq:resolution}
\sigma^2 = p_0 + p_1\cdot E + p_2\cdot E^2,
\end{equation}
where $p_0=0.155$~keV$^2$, $p_1=3.56\times10^{-4}$~keV and $p_2=1.19\times10^{-7}$.
The fitted result agrees well with the manufacturer's specification listed in Tab.~\ref{tab:detector-specification}, providing a functional form for 
the variance $\sigma$ within [20, 2700] keV range, and the basic information for gamma-peak-counts finding and correlated analysis below.

\section{Performance of the Counting Station}
\label{sec:performance}
The most important parameters of a counting station are the detection sensitivities at different energies, normally reported in the unit of Bq/kg. The sensitivities define the ability of the counting station to distinguish candidate signals from backgrounds. Lower value of sensitivity means better performance.

For a given isotope with a characteristic gamma peak with energy $E$,
the sensitivity $S(E)$ is calculated from following equation~\cite{glenn1989radiation},
\begin{equation}
S(E) = \frac {\sqrt{N^b(E,t)T/t}}{TM\eta(E)},
\label{eq:sensitivity-for-e}
\end{equation}
where $M$ is the mass of the sample to be measured,
$T$ is the time of the sample counting, $t$ is the time of background-only counting,
and $\eta(E)$ is the corresponding HPGe detection efficiency to be discussed later.
Taking into account the detector energy resolution, $N^{b}(E,t)$ is generally taken as the number of background counts collected 
within the [$E$-3$\sigma$, $E$+3$\sigma$] window\cite{1748-0221-6-08-P08010}, where the $\sigma$ is deduced from Eq.~\ref{eq:resolution}. 

The parameter $\eta(E)$ is a property related to the detector and the sample. It is mainly determined by the size of the HPGe crystal. 
The counting station has an excellent detection efficiency due to the large crystal mass, which, to our knowledge, is the largest one used currently in assaying facilities around the world. To improve the sensitivity, 
one can either increase the sample-counting time $T$ and the sample mass $M$, or suppress the background rate $N^b(E,t)/t$.

Four possible sources of backgrounds for the counting station have been studied.
First of all, high energy cosmic muons can introduce particle showers, serving as one source of the backgrounds. CJPL is the deepest underground laboratory around the world, providing excellent cosmic muon shielding. The muon flux inside the laboratory was measured to be $(2.0\pm0.4)\times10^{-10}$cm$^{-2}$s$^{-1}$\cite{Yu-Cheng:2013iaa}. So the background contribution from cosmic muons is negligible.
The second source of backgrounds comes from the radioactive isotopes inside the rocks and concrete. According to the measurement in \cite{Zeng2014}, the maximum effective radioactivities in the rock and concrete are 19.880~Bq/kg for $^{238}$U, 8.15~Bq/kg for $^{232}$Th, and 36.669~Bq/kg for $^{40}$K. With the helping of the passive shield, only a tiny fraction of these radiations can enter the counting station, thus they are also negligible. The third is the unstable isotopes in the copper and lead for the construction of the shield, which is an important limitation of the detection sensitivities.
Finally, an important background is expected to come from either the rest Rn inside the counting chamber or from the detector itself, like the electronics or the cryostat that are in close proximity.

\begin{figure}
\begin{center}
\includegraphics[width=0.99\textwidth]{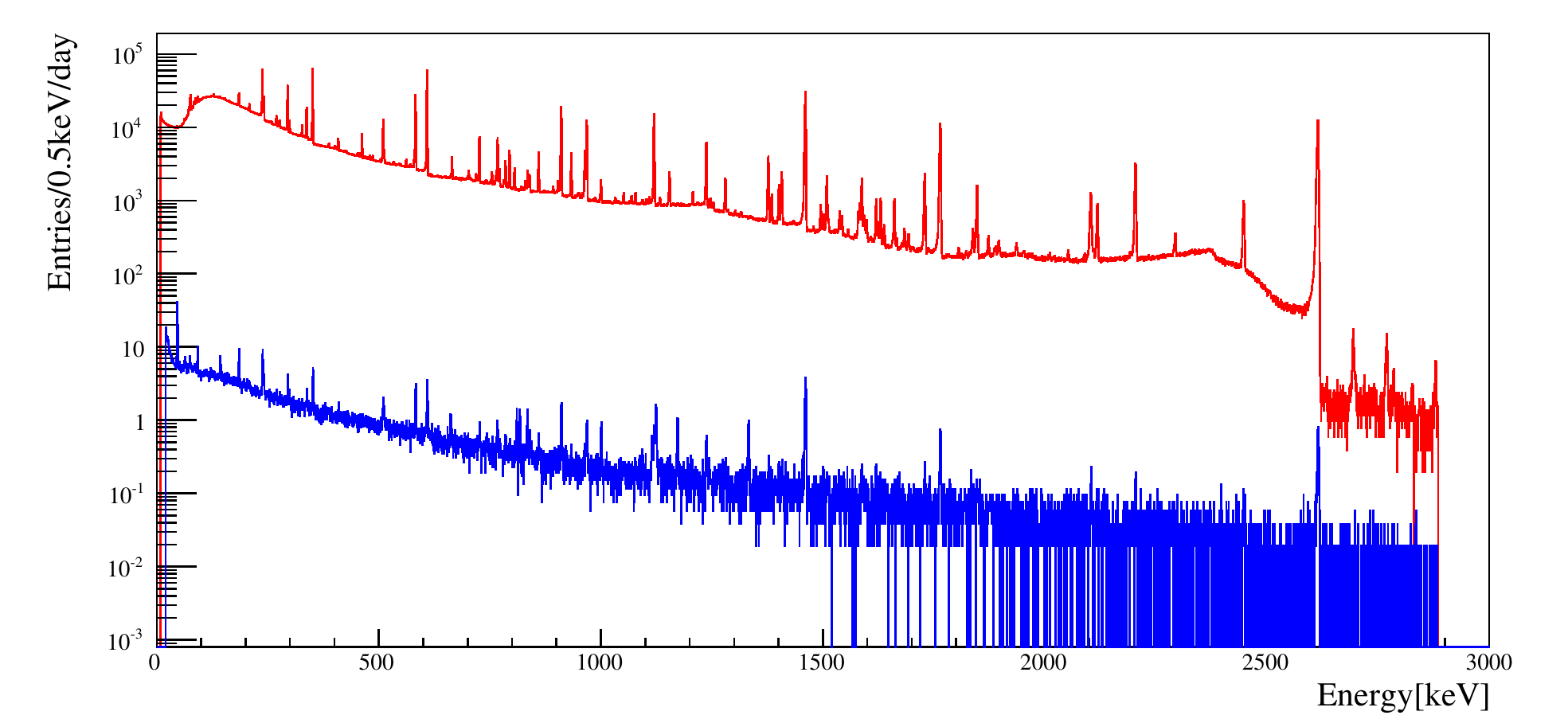}
\end{center}
\caption{Background spectra measured by the counting station with(blue)/without(red) the passive shielding system and the Rn.}
\label{fig:bkg_reduction}
\end{figure}

Fig.~\ref{fig:bkg_reduction} shows the background gamma spectrum measured by the HPGe detector with the passive shield and Rn purging system. The spectrum measured without those
is also shown for a comparison. It is seen that the shielding and Rn purging help to suppress the background level by more than 3 orders of magnitude. Specially, 
tn the energy range of 20 to 2700~keV, a background rate of $1.17\times10^4$ counts/min had been obtained without the shielding, while a much lower rate of 2.59 counts/min was acquired with the background control measures.

For comparison, the background rate $N^b(E,t)/t$ of the radioactive isotopes with characteristic energies are summarized in Tab.~\ref{tab:isotope_rate}, averaged over with
the background-only counting time $t =52.1$ days.
Typically, the sample counting time $T$ is about 7$\sim$10 days, and the sample mass $M$ is between 0.3 to 1.5 kg. 
The detection efficiency $\eta(E)$ varies from 0.2$\%$ to 10$\%$ depending on the geometrical factors and materials of samples, as well as the energy and branching ratios of the specific isotopes. Correspondingly, the sensitivities of the counting station can reach mBq/kg level for the isotopes such as K, U and Th, and even better for Co and Cs in normal operations.

\begin{table}
\centering
\begin{tabular}{lcc} \hline \hline
    Isotope/chain & Energy [keV] & Background rate [counts/day] \\ \hline
    $^{137}$Cs & 661.6 & 7.3$\pm$0.4 \\ \hline
    $^{60}$Co & 1173.2 & 6.5$\pm$0.4 \\
    $^{60}$Co & 1332.5 & 5.5$\pm$0.3 \\ \hline
    $^{40}$K & 1460.8 & 22.6$\pm$0.7 \\ \hline
    $^{235}$U & 143.8 & 32.1$\pm$0.8 \\
    $^{235}$U & 185.7 & 37.5$\pm$0.8 \\
    $^{235}$U & 205.3 & 18.2$\pm$0.6 \\ \hline
    $^{238}$U/$^{226}$Ra & 186.2 & 60.3$\pm$1.1 \\
    $^{238}$U/$^{214}$Pb & 295.2 & 30.0$\pm$0.8 \\ 
    $^{238}$U/$^{214}$Pb & 351.9 & 31.1$\pm$0.8 \\ 
    $^{238}$U/$^{214}$Bi & 609.3 & 20.9$\pm$0.6 \\ 
    $^{238}$U/$^{214}$Bi & 1120.3 & 7.8$\pm$0.4 \\ 
    $^{238}$U/$^{214}$Bi & 1764.5 & 4.5$\pm$0.3 \\ \hline
    $^{232}$Th/$^{228}$Th & 238.6 & 67.6$\pm$1.1 \\ 
    $^{232}$Th/$^{228}$Ac & 338.3 & 16.8$\pm$0.6 \\
    $^{232}$Th/$^{208}$Tl & 583.2 & 20.4$\pm$0.6 \\
    $^{232}$Th/$^{228}$Ac & 911.2 & 8.3$\pm$0.4 \\
    $^{232}$Th/$^{228}$Ac & 968.9 & 5.5$\pm$0.3 \\ 
    $^{232}$Th/$^{208}$Tl & 2614.5 & 5.6$\pm$0.3 \\ \hline  \hline
\end{tabular}
\caption{The background rates of concerned isotopes. For each energy peak, the counts are integrated within $\pm$3$\sigma$ window.}
\label{tab:isotope_rate}
\end{table}

There are many low-background HPGe counting facilities running successfully around the world~\cite{HPGeNorthAmerica}.
The properties of the representative ones are summarized in Tab.~\ref{tab:bkg_rate_exp}. 
The background rate of the PandaX counting station is comparable with those in MELISSA\cite{Finnerty:2010ar} and CORRADO\cite{Budjas:2007yj}, but higher than Gator\cite{1748-0221-6-08-P08010}and GeMPI\cite{Neder2000191,Heusser2006495}. 
The higher rate may come from the higher detection efficiency introduced by the large HPGe crystal mass, and may also come from the radioactive contaminations inside the readout electronics and cryostat.

\begin{table}
\centering
\begin{tabular}{lcp{2cm}cp{3cm}} \hline \hline
 Name & Laboratory & Crystal mass (kg) & Energy range (keV) & Background rate (counts/min) \\\hline
 MELISSA & KURF & 1.1 & 40 - 2700 & 5.42 \\\hline
 CORRADO & MPI-K & 0.93 & 100 - 2700 & $3.20\pm0.01$ \\\hline
 Gator & LNGS & 2.2 & 100 - 2700 & $0.157\pm0.001$ \\\hline
 GeMPI & LNGS & 2.15 & 100 - 2740 & $0.0279\pm0.0004$ \\\hline
 \multirow{2}{*}{PandaX} & \multirow{2}{*}{CJPL} & \multirow{2}{*}{3.69} & 100 - 2700 & 1.79\\\cline{4-5}
 & & & 20 - 2700 & 2.59\\\cline{4-5}\hline\hline
\end{tabular}
\caption{The background rates of main counting facilities around the world.}
\label{tab:bkg_rate_exp}
\end{table}

\section{Calibration and Sample Counting}
\label{sec:calib_counting}
In this section, we discuss how to calibrate our detector systems using standard either gamma sources or gammas from residual Rn gas. We will also discuss procedures and analysis in sample counting.

\subsection{Energy Calibration}
\label{sec:calib}
\begin{figure}
\centering
\includegraphics[width=0.8\textwidth]{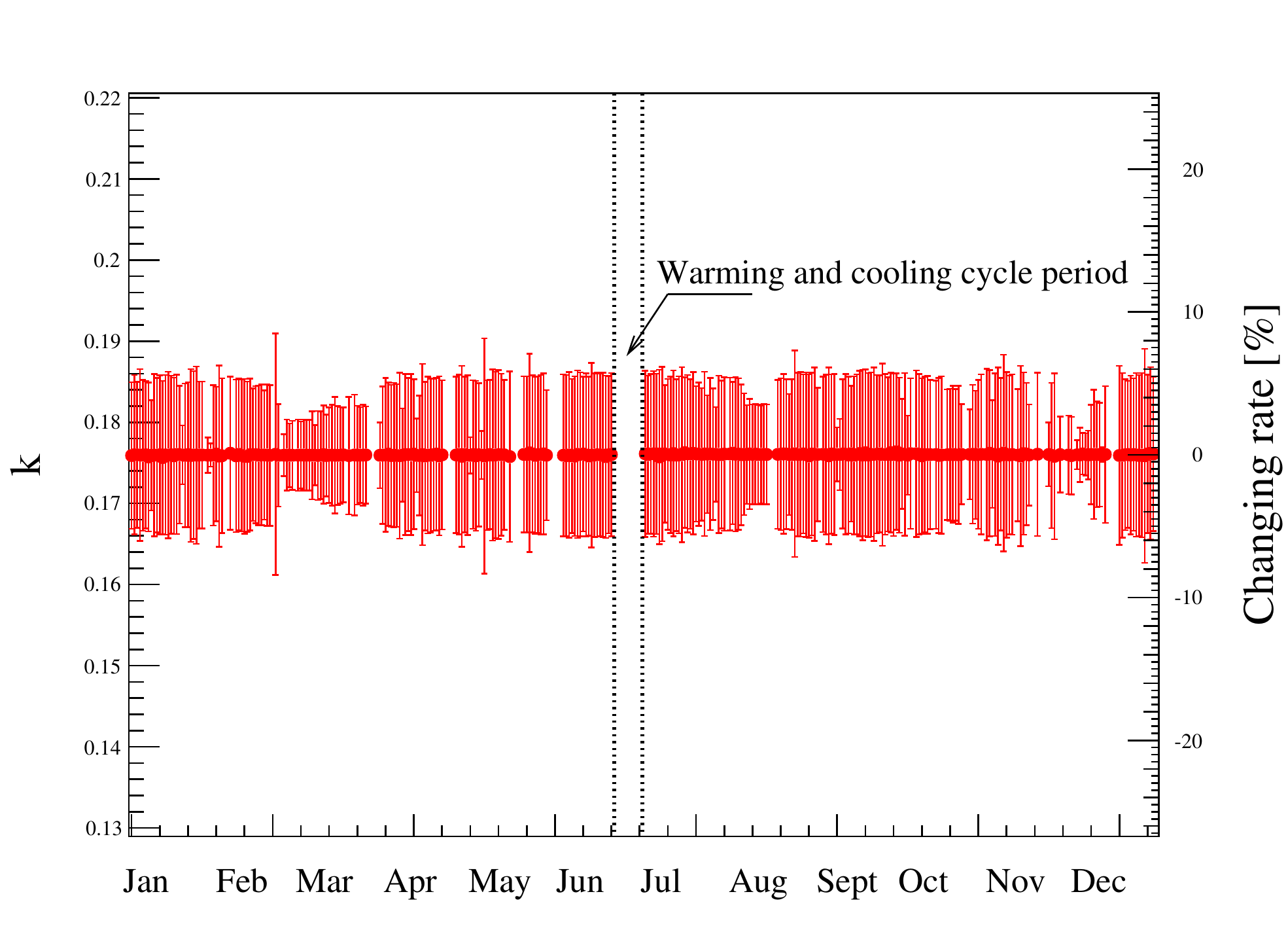}
\caption{The evolution of the energy calibration factor $k$ in 2014. The red points are the measured $k$ values, and the error bars are caused by the low count number acquired from daily data.}
\label{fig:e_calib}
\end{figure}

The detector is calibrated using the Rn gamma lines in the energy spectrum after $\sim$24 hours of data collection. Ortec DSPEC-502 is used as the data aquisition (DAQ) system. A rise time of 12 $\mu$s and flat top of 0.8 $\mu$s are used for the energy filter designed for vetoing pile-up events, which are rare in our data taking. 
Event energy, $E$, is calculated by the trapezoidal-energy-filter output of the DAQ, $h$, using the following equation,
\begin{equation}
\label{eq:e_h}
  E = kh + c,
\end{equation}
where $k$ is the calibration constant and the parameter $c$ is for possible offset.
The fluctuation of the calibration constant $k$ can be used to monitor the stability of the detector. The values of $k$ obtained in 2014 is shown in Fig.~\ref{fig:e_calib}.
Due to the low count rate, there is a large statistical uncertainty in $k$. Except for the period of warming up and cooling down cycle from June 20 to July 2, 2014 
and a few days when data storage problems happened, all the data in that year are recorded. With an average uncertainty of $\sim$6$\%$, the $k$ is kept roughly at a constant 0.176, indicating a stable status of the detector.
The data from energy calibration are also used to monitor the upper limit of Rn concentration within the detector chamber, which is stable at 1.5 Bq/m$^3$ during the normal operation. Finally, $c$ is a very small number having little effect on the energy calibration.

\subsection{Detection Efficiency Study}
\label{sec:effi}
The detection efficiency $\eta(E)$ changes with various samples due to their different geometrical structure and material composition. 
For each sample, the value is estimated with a Monte Carlo (MC) simulation program based on the Geant4 toolkit.
However, we can also calibrate the simulation with standard radiation sources. 

\begin{figure}
\centering
\includegraphics[width=0.8\textwidth]{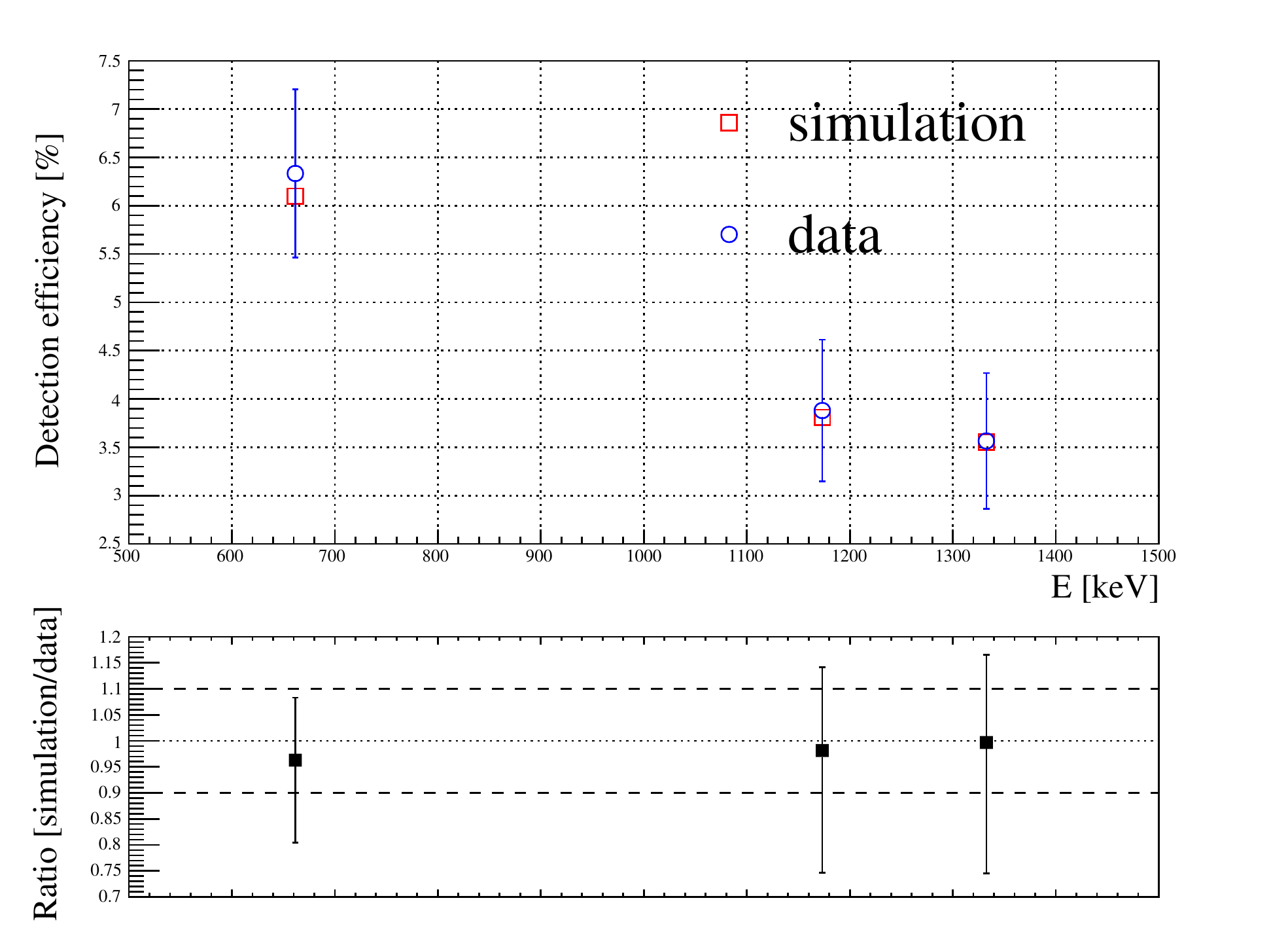}
\caption{The comparison of detection efficiencies predicted by the MC simulation program and extracted from the measured data with standard $^{137}$Cs and $^{60}$Co sources.}
\label{fig:eff_mc_data}
\end{figure}

For a given sample with the known activity $A$ of a concerned isotope as well as the sample counting time $T$, the number of total radiation events $N^{A}(T)$=$A\cdot T$ is expected.
For the gamma peak with characteristic energy $E$, the number of sample counts $N^{s}(E,T)$ in $T$ duration detected by the counting station is simulated, integrated within the [$E$-3$\sigma$, $E$+3$\sigma$] region. $N^{s}(E,T)$ consists of the net radiation counts $N^{c,s}(E,T)$ by the concerned gamma peak, and a certain amount of Compton 
background counts from higher energy gammas. To find the latter, the average value of sample counts $N^{s}_{L}(E,T)$ and $N^{s}_{R}(E,T)$ within [$E$-9$\sigma$, $E$-3$\sigma$] and [$E$+3$\sigma$, $E$+9$\sigma$] windows is taken as an approximation, assuming the Compton background has a linear slope in the whole $\pm$9$\sigma$ region.
Thus, after removing the Compton background, the net radiation counts is calculated as,
\begin{equation}
N^{c,s}(E,T) = N^{s}(E,T) - \frac{1}{2}\cdot(N^{s}_{L}(E,T) + N^{s}_{R}(E,T)),
\label{eq:netradiation-for-e}
\end{equation}
and thererefore $\eta(E)$ is calculated as
\begin{equation}
\eta(E) = \frac {N^{c,s}(E,T)}{N^{A}(T)}.
\label{eq:efficiency-for-e}
\end{equation}
The dominant uncertainty in the MC simulation comes from the relative position of the crystal with respect to the carbon fiber entrance window, or the distance of the crystal top surface to the entrance window. 
Based on the value provided by the fabricator, this distance as implemented in the MC simulation is calibrated by using two standard $^{137}$Cs and $^{60}$Co sources with known activities, as shown in Fig.~\ref{fig:eff_mc_data}.
The predictions from MC simulations agree well with the data. The maximum difference is smaller than 5$\%$.

\subsection{Actual Sample Counting and Analysis}
\label{sec:sample_counting}
Samples should be pre-treated to remove possible contaminations on the surface. In general, following procedures for the counting have been performed in PandaX:
\begin{itemize}
\item wash the sample with a type of solvent in ultrasonic for 30 - 60 minutes;
\item dry the sample with nitrogen gas;
\item seal the sample with clean plastic bag;
\item measure the weight and geometrical parameters of the sample and make a record;
\item place the sample (with bag) into the station's copper chamber;
\item start counting after 12 hours of Rn purging.
\end{itemize}

For the $i$-th characteristic gamma peak with energy $E_i$ of a concerned isotope, the characteristic activity $A_i$ is defined by
\begin{equation}
\label{eq:i_activity}
  A_i = \frac{N^{c,s}(E_i,T)/T-N^{c,b}(E_i,t)/t}{\eta(E_i)},
\end{equation}
where $N^{c,s}(E_i,T)$ and $N^{c,b}(E_i,t)$ are the number of net radiation counts at energy $E_i$ in sample counting time $T$ and background counting time $t$, respectively. The detailed calculation of $N^{c,s}(E_i,T)$ is described in Eq.~\ref{eq:netradiation-for-e}, as for $N^{c,b}(E_i,t)$ the basic principle is the same.
The activity $A$ of the isotope is the weighted average of all its $A_i$ values.
The weight for one characteristic gamma is determined by both the branching ratio and the detection efficiency.

\section{Screening Results for PandaX experiments}
\label{sec:result}
The materials used for the construction of the PandaX-I/II experiments have been screened with the counting station. The detailed counting results are presented in Tab.~\ref{tab:counting_results}. The use of the materials and the results are discussed in the following sections.

\begin{sidewaystable}
\centering
\resizebox{\linewidth}{!}{
\begin{tabular}{lllllllllllll}
\hline \hline
 & Material & Supplier & Use & Amount & Unit & $^{226}$Ra & $^{228}$Ac & $^{228}$Th & $^{235}$U & $^{137}$Cs & $^{60}$Co & $^{40}$K \\ \hline
 & \bf{PMTs} \\
 1. & 3" R11410-MOD KA0049 PMT & Hamamatsu & PandaX-II top$\&$bottom arrays, PandaX-I bottom array & 1 pc & mBq/pc & 2.40$\pm$1.22 & 5.18$\pm$2.50 & 4.71$\pm$2.09 & $<$2.70 & $<$0.72 & 2.66$\pm$0.57 & $<$17.33 \\
 2. & 3" R11410-MOD KA0051 PMT & Hamamatsu & PandaX-II top$\&$bottom arrays, PandaX-I bottom array & 1 pc & mBq/pc & 1.88$\pm$0.97 & $<$1.75 & $<$1.63 & $<$1.17 & $<$0.39 & 4.23$\pm$0.53 & 16.06$\pm$9.39 \\
 3. & 1" R8520-406 PMT & Hamamatsu & PandaX-II veto, PandaX-I top array & 7 pcs & mBq/pc & $<$0.12 & $<$0.14 & $<$0.15 & $<$0.11 & 0.17$\pm$0.04 & 0.52$\pm$0.05 & 9.94$\pm$0.98 \\
 4. & 3" PMT base & Fralock & PandaX-II top$\&$bottom arrays & 3 pcs & mBq/pc & 3.03$\pm$0.40 & $<$0.48 & 0.39$\pm$0.28 & 0.64$\pm$0.34 & $<$0.14 & $<$0.10 & $<$3.48 \\
 5. & 3" PMT base & Fralock & PandaX-I bottom array & 3 pcs & mBq/pc & 1.14$\pm$0.14 & $<$0.29 & 0.16$\pm$0.15 & 0.37$\pm$0.19 & 0.33$\pm$0.08 & $<$0.06 & $<$1.25 \\
 6. & 1" PMT base & Fralock & PandaX-II veto, PandaX-I top array & 2 pcs & mBq/pc & 0.73$\pm$0.18 & $<$0.29 & $<$0.32 & $<$0.18 & 0.35$\pm$0.09 & $<$0.07 & $<$2.44 \\ \hline
 & \bf{Metal} \\
 7. & Stainless steel 304L P03 & TISCO & PandaX-II inner vessel & 1.043 kg & mBq/kg & 1.34$\pm$1.11 & $<$1.73 & 6.09$\pm$1.89 & $<$1.68 & $<$0.47 & 0.97$\pm$0.45 & $<$8.10 \\
 8. & Stainless steel 304L P04 & TISCO & PandaX-II over flow chamber & 0.411 kg & mBq/kg & $<$1.43 & $<$2.41 & $<$1.90 & $<$2.00 & $<$0.68 & 0.75$\pm$0.63 & $<$11.63 \\
 9. & Stainless steel 304L P05-1 & CISRI & PandaX-II electrodes & 0.40 kg & mBq/kg & $<$1.28 & $<$1.93 & $<$1.53 & $<$2.35 & 0.60$\pm$0.58 & $<$0.53 & $<$13.43 \\
 10. & Stainless steel 304L P05-2 & CISRI & PandaX-II electrodes & 0.55 kg & mBq/kg & $<$1.45 & $<$1.64 & $<$1.27 & $<$1.67 & 0.49$\pm$0.45 & $<$0.40 & 9.24$\pm$8.76 \\
 11. & Stainless steel 304L P07-1 & CISRI & PandaX-II CF flange & 0.729 kg & mBq/kg & $<$1.9 & $<$3.0 & $<$3.4 & $<$2.7 & 1.4$\pm$1.0 & $<$0.7 & $<$16.2 \\
 12. & Stainless steel 304L P07-2 & CISRI & PandaX-II CF flange & 0.261 kg & mBq/kg & $<$1.70 & $<$2.74 & $<$1.71 & $<$2.43 & 2.36$\pm$0.96 & 1.03$\pm$0.75 & $<$13.95 \\
 13. & Stainless steel 316Ti & Changrong & PandaX-I inner vessel & 0.738 kg & mBq/kg & $<$1.68 & 4.22$\pm$2.71 & $<$2.17 & 4.88$\pm$2.73 & $<$0.93 & 5.95$\pm$0.83 & $<$12.80 \\
 14. & Stainless steel 316Ti & Custom & screws for PandaX-I top PMT array installation & 525 pins & mBq/pin & 0.001(3) & 0.004(6) & 0.004(3) & 0.009(2) & 0.002 & 0.004 & 0.020(9) \\
 15. & Carbon steel GB/T699-1999 & Custom & bolts for PandaX-II inner vessel sealing & 0.48 kg & mBq/kg & 19.2$\pm$7.5 & $<$9.6 & $<$5.5 & $<$8.5 & 3.6$\pm$3.1 & $<$2.5 & $<$92.8 \\
 16. & OFHC copper & CHINALCO & PandaX-I$\&$PandaX-II outer vessel, shaping rings, TPC plates & 2.836 kg & mBq/kg & $<$0.38 & $<$0.36 & $<$0.51 & $<$0.86 & $<$0.16 & 0.20$\pm$0.09 & 4.00$\pm$1.00 \\ \hline
 & \bf{Plastic} \\
 17. & PTFE & McMaster & PandaX-II TPC reflectors, PandaX-I TPC top reflectors & 0.955 kg & mBq/kg & 3.16$\pm$0.96 & $<$1.49 & $<$1.41 & $<$1.26 & 1.30$\pm$0.44 & $<$0.34 & $<$7.37 \\
 18. & PTFE & Sanxin & PandaX-I TPC side$\&$bottom reflectors & 0.129 kg & mBq/kg & $<$6.16 & $<$9.48 & $<$8.94 & 9.94$\pm$6.47 & 5.86$\pm$2.77 & $<$2.31 & 88.67$\pm$51.31 \\
 19. & Epoxy & Hanko & PandaX-II PMT insulation, PandaX-I feedthrough sealing & 0.036 kg & mBq/kg & 357$\pm$18 & 143$\pm$30 & 259$\pm$32 & 25$\pm$16 & 21$\pm$7 & $<$11 & 311$\pm$125 \\ \hline
 & \bf{Cables$\&$connectors} \\
 20. & HV cable & Matsusada & PandaX-I$\&$PandaX-II high voltage & 2.6 m & mBq/m & $<$3.20 & $<$3.57 & $<$2.69 & $<$2.74 & $<$0.89 & $<$0.70 & $<$14.77 \\
 21. & Kapton cable & Accu-Glass & PandaX-I$\&$PandaX-II signal & 22.2 m & mBq/m & $<$0.124 & $<$0.199 & $<$0.116 & $<$0.063 & $<$0.066 & $<$0.049 & $<$1.586 \\
 22. & RG316D cable & Accu-Glass & PandaX-I$\&$PandaX-II signal & 51 m & mBq/m & $<$0.078 & $<$0.179 & $<$0.071 & $<$0.042 & $<$0.037 & $<$0.025 & $<$1.001 \\
 23. & Optical fiber & Accu-Glass & PandaX-I$\&$PandaX-II photon signal calibration & 1 pc & mBq/pc & $<$2.33 & $<$1.93 & $<$2.04 & $<$3.50 & 1.47$\pm$0.60 & $<$0.57 & 16.36$\pm$10.18 \\
 24. & Socket & Accu-Glass & PandaX-I$\&$PandaX-II PMT base & 100 pins & mBq/pin & 0.002(9) & 0.005(2) & 0.002(2) & 0.005(7) & 0.003(8) & 0.001(3) & 0.049(1) \\
 25. & Socket & Kyocera & PandaX-I$\&$PandaX-II inner vessel feedthrough & 99 pins & mBq/pin & 0.003(9) & 0.006(9) & 0.004(6) & 0.008 & 0.002(1) & 0.001(8) & 0.035(6) \\
 26. & Capacitor & Kyocera & PandaX-I$\&$PandaX-II PMT base & 1000 pcs & mBq/pc & 0.15$\pm$0.04 & $<$0.10 & $<$0.04 & $<$0.14 & $<$0.03 & $<$0.02 & $<$0.31 \\
 27. & Resistor & Kyocera & PandaX-I$\&$PandaX-II electric field & 18 pcs & mBq/pc & 0.030(4) & 0.058(9) & 0.023(7) & 0.026(8) & 0.011(9) & 0.011(6) & 0.344 \\
 28. & Soldering tin 3mm & Lucas-Milhaupt & PandaX-I$\&$PandaX-II device welding & 0.454 kg & mBq/kg & 8.81$\pm$1.67 & 38.55$\pm$3.75 & 30.98$\pm$3.41 & $<$2.39 & $<$1.81 & $<$0.96 & $<$26.17 \\
 29. & Soldering tin 1mm & Lucas-Milhaupt & PandaX-I$\&$PandaX-II device welding & 0.124 kg & mBq/kg & $<$9.59 & $<$7.74 & 23.69$\pm$3.11 & $<$52.30 & 5.56$\pm$4.11 & $<$2.01 & $<$40.29 \\
\hline
\hline
\end{tabular}}
\caption{The detailed material screening results for the PandaX-I and PandaX-II dark matter experiments.}
\label{tab:counting_results}
\end{sidewaystable}

\subsection{Stainless Steel}
\label{sec:res_ss}
Stainless steel is used for the building of the inner vessels, which serve as the containers of xenon and the detector, in the PandaX dark matter experiments. The large mass and the close distance from the sensitive volume make the inner vessel one of the major sources of background events. In the PandaX-I experiment, the inner vessel consumed 525~kg of stainless steel while only 257~kg was used in PandaX-II, in which no bottom flange was built. In PandaX-I, the steel was chosen from the market with the lowest background radiation. MC simulation shows that about 1/4 of the electron-recoil (ER) background was contributed by the inner vessel in PandaX-I as shown in Tab.~\ref{tab:mc_er}. In PandaX-II, a new inner vessel was constructed with steel made specially by the China Iron and Steel Research Institute Group (CISRI) from selected raw materials, resulting in much lower background radiation. The screening results indicate that the $^{60}$Co activity in PandaX-II steel was only about 1/6 of that in PandaX-I. Stainless steel used in PandaX-II is the purest used in the xenon dark matter experiments around the world thus far\cite{Aprile:2011ru, Akerib:2014rda}.

\begin{table}
\centering
\begin{tabular}{lcc}
\hline \hline
Item  & PandaX-I [mDRU]  & PandaX-II [mDRU]    \\\hline
PMTs+bases &6.96 &0.097 \\
Inner vessel &3.62 &0.045 \\
PTFE panels &0.64 &0.021 \\
Copper outer vessel &0.87 &0.016 \\
Other IV components &1.47 &0.026 \\ \hline
Total &13.56 &0.205 \\ \hline
\hline
\end{tabular}
\caption{Radiation background contributed by different parts in the PandaX-I and PandaX-II deduced from MC simulations. Here 1 mDRU = 10$^{-3}$ evts/keV/kg/day. The background from radioactive contaminations in xenon are not listed here.}
\label{tab:mc_er}
\end{table}

\subsection{Polytetrafluoroethylene (PTFE)}
\label{sec:ptfe}
PTFE was used to build the field cage of the detector, mainly the shaping ring supporters and photon reflectors. PandaX-I used 13~kg of PTFE, and PandaX-II used 51~kg due to the larger detector and additional layer of photon reflector. The PTFE used in PandaX-II had lower level of $^{137}$Cs and $^{40}$K contamination, leading to lower ER background.

The alpha-capture process of $^{13}\text{C}+\alpha\to^{16}\text{O}+n$ and $^{19}\text{F}+\alpha\to^{22}\text{Na}+n$ produce neutrons, which are the most dangerous background in dark matter search. The $\alpha$ particles are generated by the uranium and thorium decay chain. High abundance of $^{13}$C (1.07$\%$ in natural carbon) and $^{19}$F (almost 100$\%$ in natural fluorine) in PTFE (CF$_3$-(C$_2$F$_4$)$_n$-CF$_3$) make them the main sources of neutron background in PandaX dark matter experiments. 
Lower levels of $^{235}$U, $^{238}$U and $^{232}$Th contaminations in PTFE were found in PandaX-II in comparison with PandaX-I.

\subsection{Photonmultiplier Tubes (PMTs)}
\label{sec:pmt}

The technique of dual phase xenon time projection chamber is employed in both PandaX-I and PandaX-II. Two arrays of PMTs were constructed to detect the primary and proportional light from the collisions in liquid xenon. The top PMT array of PandaX-I consisted 143 PMTs of Hamamatsu R8520-406 (1-inch), and the bottom array consisted 37 PMTs of Hamamatsu R11410-MOD (3-inch). PandaX-II used 55 R11410-MOD PMTs in the top and bottom array, respectively, and additional 48 R8520-406 PMTs for veto purpose. These models of PMTs were fabricated specifically for low radioactive background application. The R8520-406 PMT has the same level of radioactivity as those used in the XENON100 experiments\cite{Aprile:2011ru}. Considering the larger size, the R11410-MOD PMTs also showed good radiopurity. MC simulation indicates that PMTs are the main source of ER background in both PandaX-I and PandaX-II (Tab.~\ref{tab:mc_er}).

\subsection{Miscellaneous Components}
\label{sec:misc}

More than 80 samples, including copper, cables and connectors, had been screened by the gamma counting station for PandaX experiments. Radiopure OFHC copper provided by the China Luoyang Copper Co. LTD (CHINALCO) was selected to build the outer vessel and shaping rings of the detector. Clean cables and connectors were selected based on the counting results. The ER background budgets of the components estimated by MC simulation are summarized in Tab.~\ref{tab:mc_er} by applying the same data selection cuts as used in the data analysis. PandaX-II has much lower background level than PandaX-I due to its larger detector so that the self-shielding effect of liquid xenon helps to shield more external radiations.

\section{Summary and Outlook}
\label{sec:summary}
A gamma counting station had been set up in the CJPL for the background counting of the PandaX-I/II dark matter experiments. 
An excellent counting sensitivity has been achieved due to the muon-free environment, the well-designed passive shielding, and the large sensitive HPGe mass.
The counting station has screened various of materials and component samples for the construction of the PandaX infrastructure and detectors. The results are used to keep the background in accepted levels for the different stages of the PandaX experiments.

An upgrade of the counting station is planned. 
The acrylic shell will be replaced by a stainless steel shell, which is able to hold a vacuum of 0.1 Pa. 
By vacuuming the shell and filling boiling nitrogen gas, the Rn level inside the station will reach desired value much faster, improving the Rn puring efficiency.

The counting station will be working continually for the current and upcoming PandaX experiments. It will work not only for dark matter research, but also for other low background rare event searching experiments.

\section*{Acknowledgement}
\label{sec:ack}
The PandaX project has been supported by a 985-III grant from Shanghai Jiao Tong University, a 973 grant from Ministry of Science and Technology of China (No. 2010CB833005), and grants from National Science Foundation of China (Nos. 11055003, 11375114, 11435008, 11455001, 11505112 and 11525522). This work is supported in part by the Shandong University, Shanghai Key Laboratory for Particle Physics and Cosmology, Grant No. 15DZ2272100, and the CAS Center for Excellence in Particle Physics. The work has also been sponsored by Peking University and the University of Maryland. We thank the LNGS, especially Dr. Matthias Laubenstein, for helping us screen some Cu and Ti samples in the beginning.

\bibliographystyle{unsrt}
\bibliography{refs.bib}

\section*{Appendix}
\label{sec:apd}
\begin{figure}[!htp]
\centering
\includegraphics[width=0.99\textwidth]{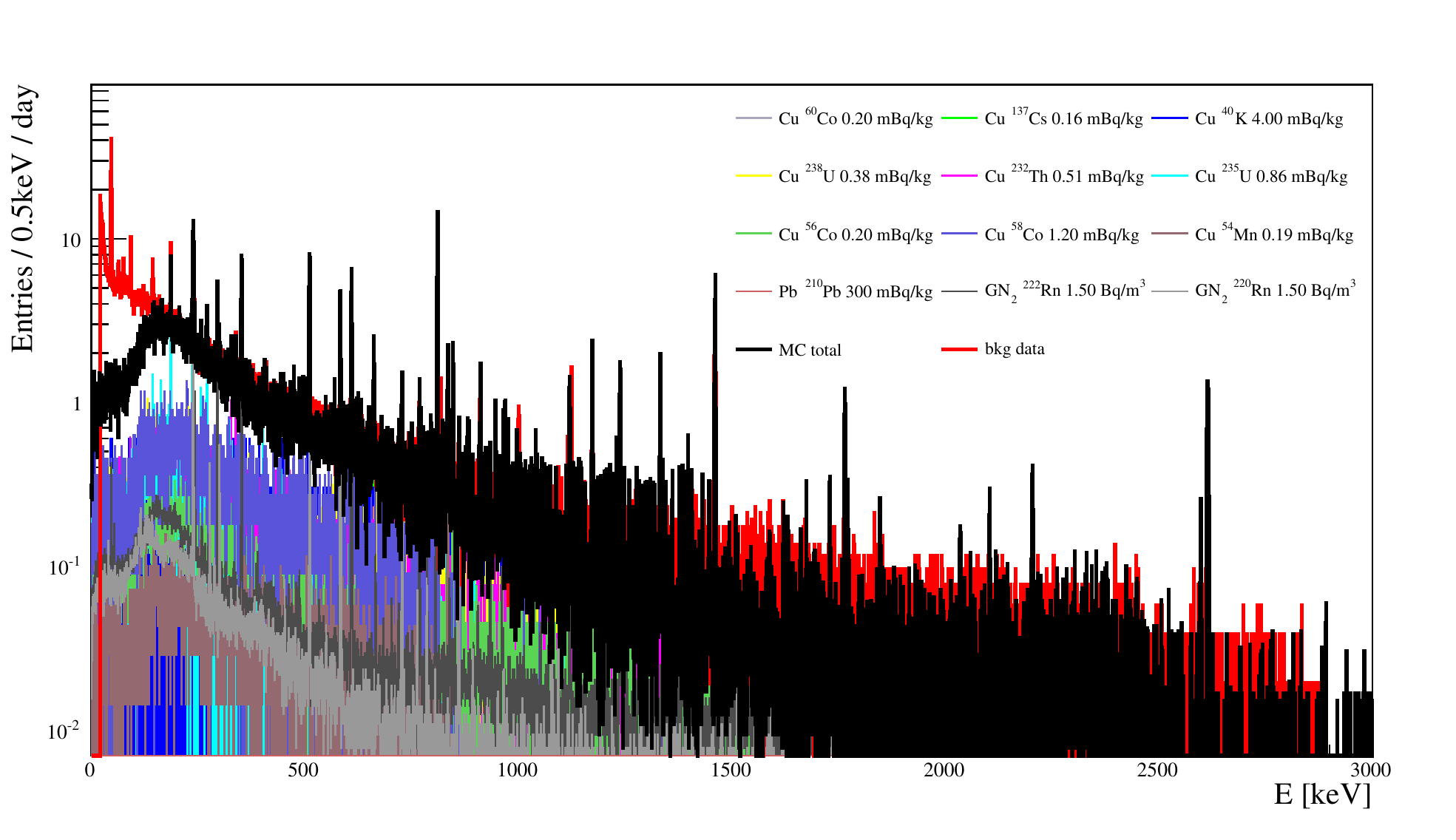}
\caption{The counting station background spectra from measurement and MC simulation.}
\label{fig:bkg_composition}
\end{figure}

In this Appendix, we use Geant4 MC to do a full simulation of the detector background from all the shielding components and Rn gas 
inside the counting chamber, as shown in the lower panel（blue) of Fig.~\ref{fig:bkg_composition}. We use the screening results listed in Tab.~\ref{tab:radiation_copper_lead} and 1.5 Bq/m$^3$ level of Rn gas as inputs, while the background from the HPGe detector itself is omitted. 
Fig.~\ref{fig:bkg_composition} shows the energy spectra from the actual measurement (red) and MC simulations (black), respectively. 
For the MC simulation, individual contributions from different isotopes are shown as curves with different colors. 
A good agreement is seen between the data and simulation within [150, 3000] keV region. The discrepancy at the low-energy region (below 150 keV) may come from the noise produced by the internal electronics and signal cables of the detector.

\end{document}